\documentclass[aps,pra,reprint,amsmath,amssymb,superscriptaddress,nofootinbib]{revtex4-2}

\usepackage{graphicx}
\usepackage{dcolumn}
\usepackage{bm}
\usepackage{hyperref}
\usepackage{braket}
\usepackage{mathtools}
\usepackage{xcolor} 
\usepackage{empheq}

\begin{document}

\title{Exact Geometric Typicality and Bipartite Entanglement\\
from the Projected Central Limit Theorem on Hyperspheres}

\author{Zhi-Wei Wang}
\email{zhiweiwang.phy@gmail.com}
\affiliation{College of Physics, Jilin University,
Changchun, 130012, People's Republic of China}
\affiliation{Computer Science, University of York, York YO10 5GH, United Kingdom}

\author{Pei-Wen Li}
\affiliation{College of Physics, Jilin University,
Changchun, 130012, People's Republic of China}

\author{Samuel L.\ Braunstein}
\email{sam.braunstein@york.ac.uk}
\affiliation{Computer Science, University of York, York YO10 5GH, United Kingdom}

\date{\today}

\begin{abstract}
{
Starting from the exact Projected Central Limit Theorem on hyperspheres, we rederive the Beta distribution for subsystem occupation probabilities and Lubkin's purity formula from elementary hyperspherical moments, quantifying the finite-size ``platykurtic'' suppression of tails relative to the Gaussian approximation used in standard eigenstate-thermalization and typicality treatments. Our main new result concerns the bipartite quantum mutual information $\langle I(A{:}B)\rangle$ for Haar-random pure states. We show that its full asymptotic expansion in $1/N$ admits a Bernoulli-factorized form in which every order $k \ge 1$ carries the symmetric factor $(d_A^{2k}-1)(d_B^{2k}-1)$ and all higher odd-order corrections vanish identically. Through an exact algebraic reorganization of Page's formula (conjectured in Ref.~\cite{Page1993} and subsequently proven~\cite{Foong1994, SanchezRuiz1995, Sen1996}), we establish that the leading finite-size correction separates into a dominant $\mathfrak{su}(d_A) \otimes \mathfrak{su}(d_B)$ bipartite quantum coherence contribution $(d_A^2 - 1)(d_B^2 - 1)/(2N)$ and a subtracted classical-probability (Cartan $\otimes$ Cartan) contribution $(d_A - 1)(d_B - 1)/(2N)$, and we trace this separation to the difference between diagonal and eigenvalue entropies via Schur's majorisation theorem, with the dimensional counts $(d-1)$ and $(d^2-1)$ acquiring meaning through the Cartan structure of the generalised Bloch decomposition. These results admit a single non-perturbative closed form: the exact typical mutual information factors as $\langle I(A{:}B)\rangle = (d_A^2-1)(d_B^2-1)\,\mathcal{G}(d_A,d_B,d_E)$, with $\mathcal{G}$ given by an explicit Bose--Einstein integral whose asymptotic expansion in $1/N$ reproduces the Bernoulli series.
}
\end{abstract}
\maketitle

\section{Introduction}

A central question in modern physics is how isolated quantum systems,
evolving unitarily, come to exhibit thermal behavior~\cite{Deutsch1991,
Srednicki1994, Rigol2008}. The resolution lies in the concepts of
quantum typicality~\cite{Goldstein2006, Popescu2006} and the
eigenstate thermalization hypothesis (ETH)~\cite{DAlessio2016}.
Conventional wisdom posits that, due to the concentration of measure
in high-dimensional Hilbert spaces, small subsystems trace out to
thermal states with fluctuations that vanish in the thermodynamic
limit.

The mathematical underpinning for these typicality concepts relies
on the concentration of measure in high-dimensional Hilbert spaces.
In mathematical physics, this concentration is traditionally bounded
by L\'evy's lemma~\cite{Popescu2006}, which provides an exponential
upper bound on the probability that a subsystem deviates from the
microcanonical average. L\'evy's lemma yields an inequality---a
worst-case bounding envelope---rather than the exact statistical
profile of the fluctuations.

To model the probability density of subsystem fluctuations, standard
treatments invoke Gaussian approximations derived from the Projected
Central Limit Theorem (PCLT): projections of high-dimensional
uniformly distributed unit vectors are approximately
Gaussian~\cite{Klartag2007}. This approximation is
accurate when the traced-out dimension is large, but has two
structural drawbacks in finite systems: the Gaussian has infinite
support and so assigns positive density to unphysical values
$P_k<0$ or $P_k>1$, and it overestimates the probability of large
deviations.

Finite-size regimes of this kind arise in Noisy
Intermediate-Scale Quantum (NISQ) devices~\cite{Preskill2018} and
mesoscopic quantum heat engines~\cite{Binder2018}, and also appear
in discussions of the black hole information paradox~\cite{Page1993,
Hayden2007}. In such systems the weight of rare fluctuations
controls device stability, error thresholds, and correlation
endpoints, so the precise tail behavior matters.

In this paper we work directly with the exact PCLT
distribution on hyperspheres~\cite{Wang2023,Wang2024}, whose radial
projection gives a Beta distribution rather than a Gaussian. Most of
the resulting single-subsystem statements---the Beta law for
diagonal elements and Lubkin's purity formula---are known from
random-matrix and induced-measure treatments~\cite{Zyczkowski2001,
Bengtsson2006, Lubkin1978}; our contribution there is methodological,
showing that they follow transparently from hyperspherical projection
without invoking Weingarten calculus~\cite{Weingarten1978, Collins2006}, and quantifying the ``platykurtic''
suppression of tails relative to the Gaussian approximation for a
concrete finite-size example. Our main new result concerns the bipartite quantum mutual information $\langle I(A{:}B)\rangle$ for Haar-random pure states. We show that its full asymptotic expansion in $1/N$ admits a Bernoulli-factorized form in which every order $k \ge 1$ carries the symmetric factor $(d_A^{2k}-1)(d_B^{2k}-1)$ and all higher odd-order corrections vanish identically. Through an exact algebraic reorganization of Page's formula (conjectured in Ref.~\cite{Page1993} and subsequently proven~\cite{Foong1994, SanchezRuiz1995, Sen1996}), we establish that the leading finite-size correction separates into a dominant $\mathfrak{su}(d_A) \otimes \mathfrak{su}(d_B)$ bipartite quantum coherence contribution $(d_A^2 - 1)(d_B^2 - 1)/(2N)$ and a subtracted classical-probability (Cartan $\otimes$ Cartan) contribution $(d_A - 1)(d_B - 1)/(2N)$, and we trace this separation to the difference between diagonal and eigenvalue entropies via Schur's majorisation theorem, with the dimensional counts $(d-1)$ and $(d^2-1)$ acquiring meaning through the Cartan structure of the generalised Bloch decomposition. 
These results admit a single non-perturbative closed form: the exact typical mutual information factors as $\langle I(A{:}B)\rangle = (d_A^2-1)(d_B^2-1)\,\mathcal{G}(d_A,d_B,d_E)$, with $\mathcal{G}$ given by an explicit Bose--Einstein integral whose asymptotic expansion in $1/N$ reproduces the Bernoulli series.

A companion Letter \cite{Wang2026} focuses on the non-perturbative closed form and the Borel-summation structure.

\section{Exact PCLT on Hyperspheres}

We analyze the projection of an $(n-1)$-sphere $S^{n-1}$ with uniform
probability measure onto an $m$-dimensional subspace. This setup is
classical, going back to Borel and Poincar\'e~\cite{Borel1906}; the
finite-dimensional (non-asymptotic) form we use is stated in
Refs.~\cite{Wang2023,Wang2024}. The radial projection law is
essentially Archimedes' hat-box observation, and the resulting Beta
distribution can also be read from Stam's projection theorem~\cite{Stam1982} and the
Diaconis--Freedman analysis~\cite{Diaconis1984} of finite-dimensional PCLT corrections.
The probability density for the projected coordinates
$X=(x_1,\ldots,x_m)$ is
\begin{equation}
\text{P}(X) =
\frac{\Gamma(n/2)}{\pi^{m/2}\Gamma((n-m)/2)}
\bigl(1-|X|^2\bigr)^{(n-m-2)/2},
\label{eq:exact_pclt}
\end{equation}
supported on the unit ball $|X|^2\le 1$, with $\langle x_i^2\rangle
=1/n$. The Gaussian PCLT corresponds to the asymptotic replacement
$(1-|X|^2)^K\approx e^{-K|X|^2}$ with $K=(n-m-2)/2$, which requires
the integrated-out dimension $n-m$ to be large, not merely $n$.

To obtain the distribution of $Y\equiv|X|^2$, we exploit the
spherical symmetry of Eq.~(\ref{eq:exact_pclt}). In hyperspherical
coordinates $d^m X=r^{m-1}\,dr\,d\Omega_m$, with $S_{m-1}
=2\pi^{m/2}/\Gamma(m/2)$ the unit-sphere area. Integrating out the
angles gives
\begin{equation}
\text{P}_r(r) =
\frac{2\Gamma(n/2)}{\Gamma(m/2)\Gamma((n-m)/2)}\,
r^{m-1}\bigl(1-r^2\bigr)^{\frac{n-m-2}{2}}.
\label{eq:radial_pdf}
\end{equation}
Changing variables to $Y=r^2$ (with $|dr/dY|=\tfrac{1}{2}Y^{-1/2}$)
and collecting powers gives
\begin{equation}
\text{P}_Y(Y) =
\frac{\Gamma(n/2)}{\Gamma(m/2)\Gamma((n-m)/2)}\,
Y^{m/2-1}\bigl(1-Y\bigr)^{\frac{n-m-2}{2}}.
\label{eq:beta_derived}
\end{equation}
This is the Beta distribution $\mathrm{Beta}(\alpha,\beta)$ with
\begin{equation}
\alpha=\frac{m}{2},\qquad \beta=\frac{n-m}{2},
\end{equation}
so that $|X|^2\sim\mathrm{Beta}(m/2,(n-m)/2)$. Using
$\langle Y\rangle=\alpha/(\alpha+\beta)$ one recovers
$\langle|X|^2\rangle=m/n$ and $\langle x_i^2\rangle=1/n$.

\section{Exact Distribution of Subsystem Probabilities}
\label{sec:quantum_application}

Consider an isolated system $U=S\otimes E$ with
$\mathcal{H}_U=\mathcal{H}_S\otimes\mathcal{H}_{E}$ and total dimension
$N=d_S d_E$. A normalized pure state $\ket{\Psi}\in\mathcal{H}_U$ is
a point on the real unit hypersphere $S^{2N-1}$; the microcanonical
ensemble is the uniform measure on this sphere, so the PCLT of the
previous section applies with embedding dimension $n=2N$.

Expanding $\ket{\Psi}=\sum_{i,j}c_{i,j}\,\ket{i}_S\otimes\ket{j}_{E}$
with $\sum_{i,j}|c_{i,j}|^2=1$, the probability that a measurement
on $S$ returns $\ket{k}_S$ is
\begin{eqnarray}
    P_k &\equiv& \bra{\Psi} \left( \ket{k}_S\bra{k} \otimes \mathbb{I}_{E} \right) \ket{\Psi} \nonumber \\
    &=& \sum_{j=1}^{d_E}|c_{k,j}|^2 = \sum_{j=1}^{d_E}\bigl(u_{k,j}^2+v_{k,j}^2\bigr),
\label{eq:Pk_sum}
\end{eqnarray}
where $c_{i,j}=u_{i,j}+iv_{i,j}$. Thus $P_k$ is the squared length
of the projection of a uniform vector on $S^{n-1}$ with $n=2N$ onto
a subspace of real dimension $m=2d_E$. Substituting into
Eq.~(\ref{eq:beta_derived}) gives $\alpha=d_E$ and $\beta=d_E(d_S-1)$,
so
\begin{equation}
\text{P}(P_k) =
\frac{\Gamma(d_S d_E}){\Gamma(d_E)\Gamma(d_E(d_S-1))}
P_k^{d_E-1}(1-P_k)^{d_E(d_S-1)-1}.
\label{eq:quantum_pdf}
\end{equation}
This shows that Haar-random pure states induce a Dirichlet distribution on the
diagonal of the reduced density matrix, with Beta
marginals. The exact geometric mapping of Haar-random pure states to the Dirichlet distribution on the probability simplex was pioneered early on in the physics literature~\cite{Lloyd1988, Wootters1990}, while standard modern derivations typically proceed through Wishart or Ginibre matrix ensembles~\cite{Zyczkowski2001, Bengtsson2006}. The geometric
derivation above offers the same result with a shorter path: once
one identifies $P_k$ with a squared projection norm on a uniformly
measured hypersphere, the Beta law is immediate, and the roles of
the traced ($d_E$) and retained ($d_S-1$) environment dimensions appear
directly as the two shape parameters.

The mean and variance follow from standard Beta identities:
\begin{eqnarray}
\langle P_k\rangle &=& \frac{1}{d_S},\\
\sigma^2_{P_k} &=& \frac{d_S-1}{d_S^2(N+1)},
\end{eqnarray}
recovering the maximally mixed reduced state $\rho_S=\mathbb{I}/d_S$
on average and the $1/N$ concentration of measure (canonical
typicality) for $N\gg 1$.

\subsection{Platykurtic suppression of tails in finite systems}
\label{sec:platykurtic_shield}

In finite-dimensional systems the difference between
the exact Beta distribution~(\ref{eq:quantum_pdf}) and its Gaussian
approximation is quantitatively important in the tails. Under ETH
and typicality assumptions, $P_k$ is commonly approximated by
$\mathcal{N}(\mu,\sigma^2)$ with $\mu=1/d_S$ and
$\sigma^2=(d_S-1)/[d_S^2(N+1)]$. For the general large-environment limit, the variance scales as $\sigma^2 \approx (d_S-1)/(d_S^3 d_E)$, which for the qubit case ($d_S=2$) evaluates to $\sigma^2\approx 1/(8 d_E)$. The Gaussian has zero excess
kurtosis and unbounded support, whereas the exact distribution is
bounded on $[0,1]${, and for comparable subsystem and environment dimensions, it is strictly platykurtic (excess kurtosis $<0$). While the distribution can approach a leptokurtic Gamma distribution (excess kurtosis $>0$) in the large-subsystem limit ($d_S \gg 1$ with $d_E$ fixed), the central physical mechanism for the shield remains the strict truncation of extreme right tails due to the bounded support.}

\begin{figure}[htbp]
    \centering
    \includegraphics[width=\columnwidth]{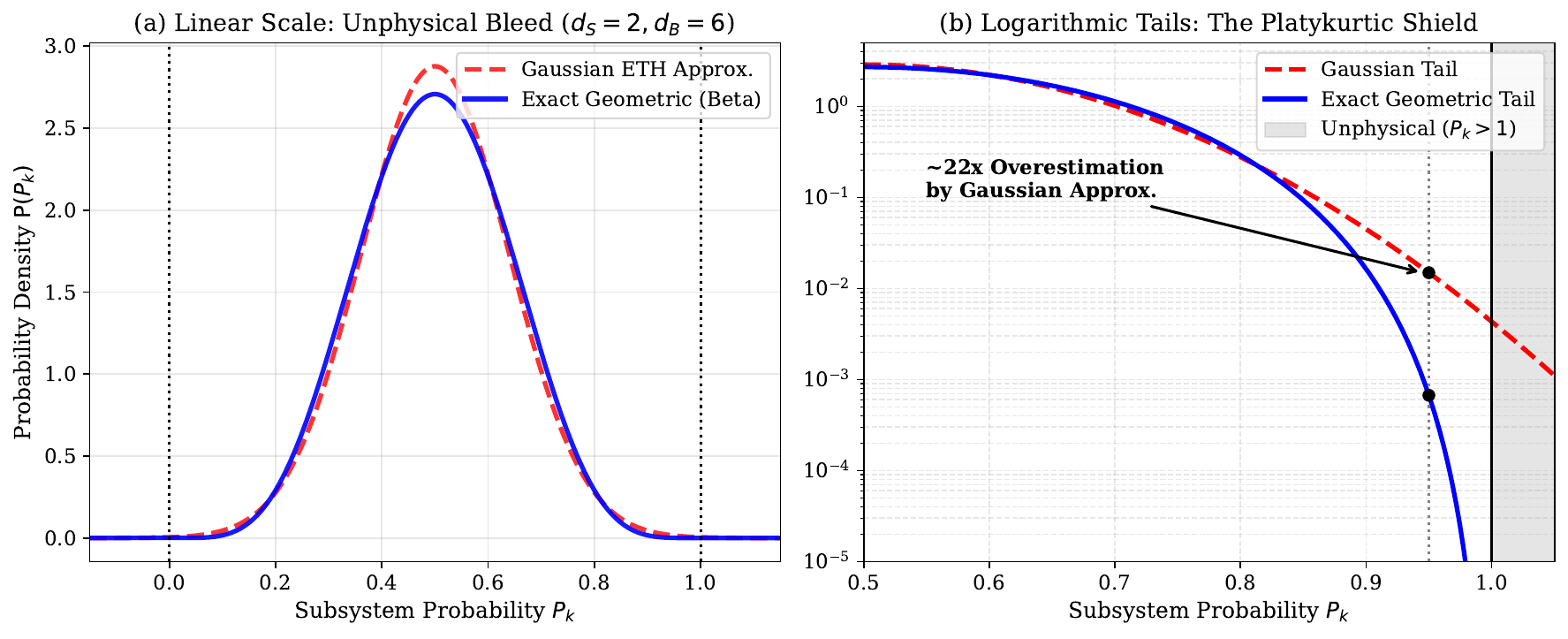}
    \caption{Geometric suppression of thermal fluctuations in a finite quantum system ($d_S=2, d_E=6$). \textbf{(a)} On a linear scale, the exact Beta distribution rigorously bounds physical probabilities to $P_k \in [0,1]$, whereas the standard Gaussian assumption erroneously bleeds into unphysical regimes (shaded red). (b) Logarithmic scale: the exact density falls below
    the Gaussian approximation by more than an order of magnitude
    near the kinematic boundary, {and the exact cumulative probability falls below by more than two orders of magnitude for $P_k>0.95$.}}
    \label{fig:tails}
\end{figure}

As a concrete example, take a single qubit ($d_S=2$) coupled to a
small environment ($d_E=6$), so $N=12$ and $P_k\sim\mathrm{Beta}(6,6)$, with
$\mu=1/2$ and $\sigma^2=1/52\approx 0.0192$. Near the mean the
Gaussian is a reasonable approximation; in the tails it is not.
At $P_k=0.95$ the exact density is
$\approx 6.7\times 10^{-4}$, while the matched Gaussian gives
$\approx 1.5\times 10^{-2}$, a factor of about $22$ too large. More critically for predicting rare physical events or modeling error thresholds, the cumulative tail probability reveals an even starker divergence. The integral of the Gaussian tail for $P_k > 0.95$ yields $\approx 5.87 \times 10^{-4}$, whereas the exact Beta tail probability is only $\approx 5.80 \times 10^{-6}$. The Gaussian approximation therefore overestimates the likelihood of this extreme statistical fluctuation by a factor of over $100$. At $P_k=1.0$, where the kinematic boundary forces the exact density to
vanish, the Gaussian still assigns density
$\approx 4.3\times 10^{-3}$, and assigns nonzero density to the
unphysical range $P_k>1$. This failure of Gaussian approximations in the extreme tails is {consistent} with Large Deviation Theory bounds, which formally demonstrate how finite kinematic boundaries strictly suppress the probability of rare typical entanglement fluctuations~\cite{Nadal2010}. Figure~\ref{fig:tails} illustrates this.

We refer to this finite-size suppression as a
``platykurtic shield'' for brevity in what follows. The point is
pragmatic rather than grand: for NISQ architectures and mesoscopic
heat engines with a genuinely small environment, a Gaussian tail estimate
can misstate the frequency of rare localization events by orders of magnitude, whereas the Beta tail follows directly from
the kinematic boundary of Hilbert space.

\section{Purity and Mutual Information from Geometric Moments}

The diagonal statistics above describe classical
measurement probabilities. To reach entanglement measures we need
off-diagonal elements of $\rho_S$ as well. Below we compute
$\langle\mathrm{Tr}(\rho_S^2)\rangle$ from hyperspherical
cross-moments, recovering Lubkin's formula, and then obtain the
leading mutual information. The off-diagonal step uses $U(1)$
phase invariance of the complex hyperspherical measure, which is
the simplest case of a Weingarten contraction~\cite{Weingarten1978, Collins2006}; we make this
explicit below.

\subsection{Subsystem purity}

The purity splits into diagonal and off-diagonal parts:
\begin{equation}
\langle\mathrm{Tr}(\rho_S^2)\rangle =
\sum_{k}\langle\rho_{kk}^2\rangle
+\sum_{k\ne l}\langle|\rho_{kl}|^2\rangle.
\end{equation}

\emph{Diagonal.}\ From $P_k\sim\mathrm{Beta}(d_E,N-d_E)$,
$\langle P_k^2\rangle=d_E(d_E+1)/[N(N+1)]$, so
\begin{equation}
\sum_{k=1}^{d_S}\langle\rho_{kk}^2\rangle
= \frac{d_E+1}{N+1}.
\label{eq:diag_purity}
\end{equation}

\emph{Off-diagonal.}\ For $k\ne l$,
$\rho_{kl}=\sum_j c_{k,j}c_{l,j}^*$, so
\begin{equation}
\langle|\rho_{kl}|^2\rangle
= \sum_{j,m}\langle c_{k,j}c_{l,j}^* c_{k,m}^* c_{l,m}\rangle.
\end{equation}
The uniform measure on the complex hypersphere is
invariant under independent $U(1)$ rotations of each component
$c_{i,j}\to e^{i\theta_{ij}}c_{i,j}$. For $k\ne l$ the expectation
above picks up a factor
$e^{i(\theta_{k,j}-\theta_{l,j}-\theta_{k,m}+\theta_{l,m})}$ under
this action, which averages to zero unless $j=m$. (This is the $n=2$
case of a Weingarten contraction~\cite{Weingarten1978, Collins2006}; what the geometric picture buys us
is that no further combinatorics is needed at this order.) With
$j=m$,
\begin{equation}
\langle|\rho_{kl}|^2\rangle
= \sum_{j=1}^{d_E}\langle|c_{k,j}|^2|c_{l,j}|^2\rangle.
\end{equation}
The squared moduli $|c_{i,j}|^2$ are the squared
coordinates of a real uniform unit vector on $S^{2N-1}$, grouped
pairwise. Their joint law is the symmetric Dirichlet distribution
$\mathrm{Dir}(1,\ldots,1)$ on the {$(N-1)$-simplex}, which gives the
cross-moment $\langle|c_{k,j}|^2|c_{l,j}|^2\rangle=1/[N(N+1)]$ for
distinct indices. Hence
\begin{equation}
\sum_{k\ne l}\langle|\rho_{kl}|^2\rangle
= \frac{d_S-1}{N+1}.
\label{eq:off_diag_purity}
\end{equation}

Adding Eqs.~(\ref{eq:diag_purity}) and (\ref{eq:off_diag_purity})
recovers Lubkin's result~\cite{Lubkin1978},
\begin{equation}
\langle\mathrm{Tr}(\rho_S^2)\rangle
= \frac{d_S+d_E}{N+1},
\end{equation}
from hyperspherical cross-moments alone. 
{While the independent $U(1)$ phase-averaging captures the purity ($n=2$ moment) without complex combinatorics, we note that calculating higher moments (e.g., $\mathrm{Tr}(\rho_S^3)$) breaks this simplicity and requires the full machinery of Weingarten calculus.}

\subsection{Mutual information at leading order}

For $N\gg 1$ the concentration of measure lets us write
$\rho_S=\mathbb{I}/d_S+\Delta$ with $\mathrm{Tr}\,\Delta=0$, and a
second-order expansion of von Neumann entropy $-\mathrm{Tr}(\rho_S\ln\rho_S)$ gives
\begin{equation}
S_{\mathrm{vN}}\approx\ln d_S-\frac{d_S}{2}\,\mathrm{Tr}(\Delta^2),
\end{equation}
where $\Delta$ is the traceless deviation from the maximally mixed state. With $\mathrm{Tr}(\Delta^2)=\mathrm{Tr}(\rho_S^2)-1/d_S$ and the
purity above,
\begin{equation}
\langle S_{\mathrm{vN}}\rangle
\approx \ln d_S-\frac{d_S^2-1}{2(N+1)}
\approx \ln d_S-\frac{d_S^2-1}{2N}
\label{Pageapproximation}
\end{equation}
i.e.\ the leading term of Page's correction~\cite{Page1993, Foong1994, Sen1996}. For a
tripartite partition $N=d_A d_B d_E$ where the composite subsystem is smaller than or equal to the traced-out environment ($d_A d_B \le d_E$), the volume-law pieces cancel and
\begin{eqnarray}
\langle I(A{:}B)\rangle  &\equiv& S_{\text{vN}}(A) + S_{\text{vN}}(B) - S_{\text{vN}}(AB) \nonumber \\
&\approx& \frac{(d_A^2-1)(d_B^2-1)}{2N}.
\end{eqnarray}
This dimensional restriction ensures that the exact formula~\cite{Page1993, Foong1994, Sen1996} does not swap its subsystem-environment arguments when evaluating $S(AB)$, preserving the algebraic factorization that follows. The factor $(d^2-1)$ is the dimension of the adjoint representation
of $SU(d)$; without the off-diagonal contribution one would instead
get $(d_A-1)(d_B-1)/(2N)$, which counts only classical diagonal
degrees of freedom.

\section{Bernoulli Factorization of the Asymptotic Expansion}

{
The Bernoulli factorisation we present in this section is the
asymptotic shadow of an exact algebraic identity hiding inside Page's
formula, which we will derive in Sec.~VI. The asymptotic form, which
foregrounds the symmetric Casimir structure $(d_A^{2k}-1)(d_B^{2k}-1)$
at every order, is sufficiently striking on its own to merit
separate presentation.
}

The leading $1/N$ result above is well known. The
novel contribution of this section is the structure of the full
asymptotic series in $1/N$.

The cleanest starting point is the exact formula for the average von Neumann entropy of a random pure state, which was originally conjectured by Page~\cite{Page1993} based on analytical limits and numerical checks, and subsequently proven rigorously~\cite{Foong1994, SanchezRuiz1995, Sen1996}
\begin{equation}
\langle S_{\mathrm{vN}}(S)\rangle
= \psi(N+1)-\psi(N/d_S+1)-\frac{d_S-1}{2(N/d_S)},
\end{equation}
where $\psi(z)$ is the digamma function.
Using the asymptotic expansion of the digamma function for
$z\gg 1$---an approach pioneered by Sen~\cite{Sen1996} to express single-subsystem typical entropies via Bernoulli numbers---one finds
\begin{equation}
\psi(z+1)\sim\ln z+\frac{1}{2z}-\sum_{k=1}^{\infty}
\frac{B_{2k}}{2k\,z^{2k}},
\end{equation}
with $B_{2k}$ the Bernoulli numbers ($B_2=\tfrac{1}{6}$,
$B_4=-\tfrac{1}{30}$, $B_6=\tfrac{1}{42}$, \ldots), which gives
\begin{equation}
    \langle S_{\mathrm{vN}}(S)\rangle
= \ln d_S-\frac{d_S^2-1}{2N} +\sum_{k=1}^{\infty}\frac{B_{2k}}{2kN^{2k}}
 \bigl(d_S^{2k}-1\bigr).
\end{equation}
The volume law and the $1/N$ correction reproduce the leading
result above; all higher-order corrections are packaged into a
single sum over even Bernoulli numbers.

Applying this to the tripartite mutual information $\langle
I(A{:}B)\rangle=\langle S(A)\rangle+\langle S(B)\rangle-\langle
S(AB)\rangle$ (maintaining the strict assumption $d_A d_B \le d_E$) and using the identity
\begin{equation}
(d_A^{2k}-1)+(d_B^{2k}-1)-(d_A^{2k}d_B^{2k}-1)
\equiv -(d_A^{2k}-1)(d_B^{2k}-1),
\end{equation}
collapses the expansion term-by-term:
\begin{eqnarray}
\langle I(A{:}B)\rangle
&=& \frac{(d_A^2-1)(d_B^2-1)}{2N}\nonumber\\
 && -\sum_{k=1}^{\infty}\frac{B_{2k}}{2kN^{2k}}
 (d_A^{2k}-1)(d_B^{2k}-1).
\label{eq:mutual_info_expansion}
\end{eqnarray}
Writing out the first several Bernoulli numbers yields
\begin{eqnarray}
\langle I(A{:}B)\rangle
&=& \frac{(d_A^2-1)(d_B^2-1)}{2N}
     -\frac{(d_A^2-1)(d_B^2-1)}{12N^2}\nonumber\\
   && +\frac{(d_A^4-1)(d_B^4-1)}{120N^4}
     -\frac{(d_A^6-1)(d_B^6-1)}{252N^6}\nonumber\\
   && +\frac{(d_A^8-1)(d_B^8-1)}{240N^8}
     -\frac{(d_A^{10}-1)(d_B^{10}-1)}{132N^{10}}\nonumber\\
   && +\mathcal{O}\bigl(1/N^{12}\bigr).
\label{eq:mutual_info_explicit}
\end{eqnarray}

\subsection{Structural features of the expansion}
\label{sec:structural_features}

Four features of Eq.~(\ref{eq:mutual_info_expansion}) are worth
noting.

\emph{(i) Symmetric factorization.}\ At every order
$k\ge 1$ the dimensional dependence factorizes as
$(d_A^{2k}-1)(d_B^{2k}-1)$. No mixed terms such as $d_A^2 d_B^4$
appear; if either subsystem is trivial ($d=1$), the whole
expansion collapses to zero. The factor $(d^2-1)$ at $k=1$ is the
$\mathfrak{su}(d)$ generator count, and the higher $(d^{2k}-1)$ factors arise
naturally from traces of Casimir-like polynomials.

\emph{(ii) Absence of higher odd-order corrections.}\ After
the leading $\mathcal{O}(1/N)$ term, the expansion contains only
even inverse powers of $N$: all $1/N^3$, $1/N^5$, $\ldots$
contributions vanish identically, a direct consequence of the
digamma asymptotics.

{
\emph{(iii) Alternating signs.}\  Since successive even Bernoulli
numbers alternate in sign, the series has signature
$(+, -, +, -, \ldots)$, so successive corrections partially cancel.  This alternating structure is a consequence of the
digamma asymptotics and is mathematically distinct from the
platykurtic (negative excess kurtosis) suppression of single-outcome
probability tails discussed in
Sec.~\ref{sec:platykurtic_shield}: the platykurtic shield concerns
the probability density $\mathrm{P}(P_k)$ of individual diagonal matrix
elements, whereas the Bernoulli series concerns the Haar-averaged
mutual information, which is an entropy functional of the full
eigenvalue spectrum.  Both phenomena, however, share a common
geometric origin in the compactness of the Hilbert-space hypersphere
$S^{2N-1}$.  The bounded support of all distributions on this sphere
constrains the signs of higher-order corrections: for the Beta
distribution this manifests as negative excess kurtosis (lighter tails
than a Gaussian), while for the mutual information it manifests as
partial cancellation of successive finite-size terms.  A precise
quantitative link between the two would require expressing the entropy
functional in terms of the full cumulant hierarchy of the Beta
distribution, which we do not pursue here.
}

\emph{(iv) Degeneracy of the first two orders.}\ The
$\mathcal{O}(1/N)$ and $\mathcal{O}(1/N^2)$ terms share the same
prefactor $(d_A^2-1)(d_B^2-1)$: the geometric leading order and the
first Bernoulli correction both track the bipartite $\mathfrak{su}(d_A) \otimes \mathfrak{su}(d_B)$ generator
dimension before higher Casimirs enter at $\mathcal{O}(1/N^4)$.

\section{Exact Algebraic Reorganization of Page's Formula}

The previous section gave the mutual information as
an asymptotic series. However, the algebraic split identified above is not inherently reliant on an asymptotic expansion or an infinite series resummation; it is an exact polynomial identity hiding inside Page's formula~\cite{Page1993, Foong1994, SanchezRuiz1995, Sen1996}. {{Note that replacing $N+1$ with $N$ in Eq.~(\ref{Pageapproximation}) is an approximation used here for the leading-order asymptotic derivation. However, in the exact rational reorganization of Page's formula presented in this section, the algebraic split natively absorbs the exact dimensions without this approximation.}} 

To see this, consider the exact rational $\mathcal{O}(1/N)$ correction terms from the evaluation for $\langle I(A{:}B)\rangle = \langle S(A)\rangle + \langle S(B)\rangle - \langle S(AB)\rangle$. When applying the $-\frac{m-1}{2n}$ correction for subsystems of dimension $m$ in an environment $n$ (assuming $d_A d_B \le d_E$), the rational terms assemble exactly to
\begin{equation}
\Delta_{\mathrm{rational}} = - \frac{d_A(d_A-1)}{2N} - \frac{d_B(d_B-1)}{2N} + \frac{d_Ad_B(d_Ad_B-1)}{2N}.
\end{equation}
Expanding these fractions and refactoring yields the exact fractional structure
\begin{eqnarray}
\Delta_{\mathrm{rational}}
&=& \frac{d_A^2 d_B^2 - d_A^2 - d_B^2 + 1}{2N} - \frac{d_A d_B - d_A - d_B + 1}{2N}\nonumber\\
&=& \frac{(d_A^2-1)(d_B^2-1)}{2N} - \frac{(d_A-1)(d_B-1)}{2N}.
\end{eqnarray}

Recombining this exact fractional identity with the remaining exact digamma functions produces a reorganization of the full mutual information
\begin{eqnarray}
\langle I(A{:}B)\rangle
&=& \psi(N+1)-\psi(N/d_A+1)-\psi(N/d_B+1)\nonumber\\
 && +\psi(N/(d_Ad_B)+1)\nonumber\\
 && +\frac{(d_A^2-1)(d_B^2-1)}{2N}
     -\frac{(d_A-1)(d_B-1)}{2N}.
\label{eq:exact_analytic_MI}
\end{eqnarray}
For integer arguments, $\psi(n+1)=H_n-\gamma$ with $H_n
=\sum_{j=1}^{n}1/j$, and the Euler--Mascheroni constants cancel
pairwise:
\begin{eqnarray}
\langle I(A{:}B)\rangle
&=& \bigl[H_{N}-H_{N/d_A}-H_{N/d_B}
      +H_{N/(d_Ad_B)}\bigr]\nonumber\\
    && +\frac{(d_A^2-1)(d_B^2-1)-(d_A-1)(d_B-1)}{2N}. \nonumber\\
    &&
\label{eq:exact_analytic_MI_harmonic}
\end{eqnarray}

The harmonic-number combination in
Eq.~(\ref{eq:exact_analytic_MI_harmonic}) evaluates exactly to Page's formula~\cite{Page1993, Foong1994, SanchezRuiz1995, Sen1996} as a function of $(d_A,d_B,d_E)$; we do not
claim the value of $\langle I(A{:}B)\rangle$ as new. What is new
is the exact algebraic split visible in the second line: the finite-size
piece separates cleanly into a purely bipartite $\mathfrak{su}(d_A) \otimes \mathfrak{su}(d_B)$ quantum correlation contribution
$(d_A^2-1)(d_B^2-1)$ and a subtracted classical-probability
contribution $(d_A-1)(d_B-1)$. The Bernoulli series presented in Section~V illustrates why the subsequent digamma residuals strictly preserve this symmetric geometric factorization at all higher even orders.

{
\subsection{Interpretation: Diagonal Entropy, Eigenvalue Entropy, and the Cartan Subtraction}
\label{sec:interpretation}

The algebraic difference
\begin{equation}
\Delta_{\mathrm{alg}}
= \frac{(d_A^2-1)(d_B^2-1)}{2N}
 -\frac{(d_A-1)(d_B-1)}{2N}
\label{eq:Delta_alg}
\end{equation}
is not merely a
numerical coincidence between Lie-algebraic dimensional counts.  It
reflects an exact decomposition of the typical mutual information into
a \emph{diagonal} (classical-probability) contribution and an
\emph{eigenvalue correction} (quantum-coherence) contribution, rooted
in the structure of Page's formula~\cite{Page1993, Foong1994, SanchezRuiz1995, Sen1996} and Schur's majorization theorem~\cite{Schur1923, Marshall2011, Horn2012}.

The exact formula for the average von Neumann entropy of a random
pure state can be written as~\cite{Page1993, Foong1994, SanchezRuiz1995, Sen1996}
\begin{equation}
\langle S_{\mathrm{vN}}(S)\rangle
= \bigl[\psi(N+1) - \psi(d_E+1)\bigr]
  - \frac{d_S - 1}{2d_E}\,,
\label{eq:Page_split}
\end{equation}
where $d_S$ and $d_E$ represent the subsystem and environment dimensions respectively, and
$N = d_S d_E$.  The two pieces of this formula have distinct physical origins.

The bracketed term represents the average diagonal entropy~ \footnote{{The term ``diagonal entropy" was introduced to refer to the entropy of the diagonal of $\rho$ in the energy basis, used as a measure of equilibration~\cite{Polkovnikov2011}. The
identification of  $\psi(N+1)-\psi(d_E+1)$ with the average entropy of a symmetric Dirichlet distribution predates that work~\cite{Lloyd1988, Wootters1990}.}}, defined as
\begin{eqnarray}
\langle S_{\mathrm{diag}}(S)\rangle
&\equiv&
\Bigl\langle -\sum_{k=1}^{d_S} P_k \ln P_k \Bigr\rangle\nonumber \\
&=& \psi(N+1) - \psi(d_E+1)\,,
\label{eq:S_diag}
\end{eqnarray}
where $P_k = \rho_{kk}$ are the diagonal matrix elements evaluated in a fixed
basis. {As derived in Sec.~\ref{sec:quantum_application}}, this identity follows from the Dirichlet distribution of the
diagonal elements~\cite{Zyczkowski2001, Lloyd1988, Wootters1990}: $\{P_k\} \sim \mathrm{Dir}(d_E, \ldots, d_E)$ on the
$(d_S{-}1)$-simplex.  For any individual component $P_k$ drawn from a symmetric
Dirichlet distribution with concentration parameter $\alpha = d_E$ and
total parameter $\alpha_0 = d_S d_E = N$, we have the expectation value
\begin{eqnarray}
\langle P_k \ln P_k \rangle
&=& \frac{\alpha}{\alpha_0}\bigl[\psi(\alpha+1) - \psi(\alpha_0+1)\bigr]\nonumber \\
&=& \frac{1}{d_S}\bigl[\psi(d_E+1) - \psi(N+1)\bigr],
\end{eqnarray}
and summing this term over all $k$ yields Eq.~(\ref{eq:S_diag}).

The second piece, $-(d_S-1)/(2d_E)$, is the eigenvalue correction. Because the von Neumann entropy is
computed from the eigenvalues $\{\lambda_k\}$ of $\rho_S$ rather
than from its diagonal elements $\{P_k\}$, and because Schur's theorem~\cite{Schur1923, Marshall2011, Horn2012} establishes that
the eigenvalue vector majorizes the diagonal vector, the entropy of
the eigenvalues is rigorously bounded above by the entropy of the diagonal, yielding $S_{\mathrm{vN}} \le S_{\mathrm{diag}}$~\cite{Wehrl1978}.  The average deficit between the two is
precisely $(d_S - 1)/(2d_E)$.  

{
The environment dimension $d_E$ plays a unifying role: it controls both the
concentration of $P_k$ around the simplex centre (through the Beta
shape parameter $d_E$, producing the platykurtic shield of
Sec.~\ref{sec:platykurtic_shield}) and the magnitude of the eigenvalue penalty
$(d_S-1)/(2d_E)$.  These are distinct effects with different
functional forms (the boundary suppression is exponential in $d_E$,
whereas the eigenvalue penalty is polynomial), but they share a
common origin in the concentration of measure on $S^{2N-1}$ that the
parameter $d_E$ controls.  In particular, the mathematical machinery
that produces the Beta distribution for diagonal probabilities is the
same that fixes the structure of Page's formula for eigenvalue
entropies; the parameter $d_E$ is the bridge between the two.
}

Note that the numerator $(d_S - 1)$ is the dimension of the probability simplex on which the diagonal elements live, which is also precisely the dimension of the Cartan subalgebra of $\mathfrak{su}(d_S)$~\cite{Georgi1999}. This is not a coincidence: the Cartan generators are exactly the traceless diagonal Hermitian matrices, and $(d_S - 1)$ counts the independent diagonal degrees of freedom constrained by the normalization $\sum_k P_k = 1$.

Applying the split from Eq.~(\ref{eq:Page_split}) to each entropy term in the mutual information expression
$\langle I(A{:}B)\rangle = \langle S(A)\rangle + \langle S(B)\rangle
- \langle S(AB)\rangle$ (maintaining the dimensional assumption $d_A d_B \le d_E$ throughout) gives
\begin{eqnarray}
\langle I(A{:}B)\rangle
&=& \langle I_{\mathrm{diag}}(A{:}B)\rangle
+ \Delta_{\mathrm{eigenvalue}}\,,
\label{eq:MI_decomp}
\end{eqnarray}
where $\langle I_{\mathrm{diag}}(A{:}B)\rangle$ is the classical-probability (diagonal) contribution, and $\Delta_{\mathrm{eigenvalue}}$ is the quantum-coherence (eigenvalue correction) contribution. The diagonal mutual information evaluates to
\begin{eqnarray}
\langle I_{\mathrm{diag}}(A{:}B)\rangle
&=& \langle S_{\mathrm{diag}}(A)\rangle
 + \langle S_{\mathrm{diag}}(B)\rangle
 - \langle S_{\mathrm{diag}}(AB)\rangle \nonumber\\
&=& \psi(N{+}1) - \psi(N/d_A{+}1)
 - \psi(N/d_B{+}1) \nonumber\\
&&+ \psi(N/(d_Ad_B){+}1) \nonumber\\
&=& H_N - H_{N/d_A} - H_{N/d_B} + H_{N/(d_Ad_B)}\,, \nonumber\\
&&
\label{eq:I_diag}
\end{eqnarray}
and the eigenvalue correction assembles to
\begin{eqnarray}
\Delta_{\mathrm{eigenvalue}}
&=& -\frac{d_A-1}{2(N/d_A)}
   -\frac{d_B-1}{2(N/d_B)}
   +\frac{d_Ad_B-1}{2(N/(d_Ad_B))} \nonumber\\[4pt]
&=& \frac{(d_A^2{-}1)(d_B^2{-}1) - (d_A{-}1)(d_B{-}1)}{2N}\,.
\label{eq:Delta_ev}
\end{eqnarray}

At leading order in $1/N$, expanding the digamma functions gives
$\langle I_{\mathrm{diag}}\rangle \sim (d_A-1)(d_B-1)/(2N)$, which allows us to write
\begin{eqnarray}
\langle I(A{:}B)\rangle
&\sim&
\frac{(d_A-1)(d_B-1)}{2N}
\nonumber\\
&&+
\frac{(d_A^2{-}1)(d_B^2{-}1) - (d_A{-}1)(d_B{-}1)}{2N}\nonumber \\
&=&
\frac{(d_A^2{-}1)(d_B^2{-}1)}{2N}\,,
\end{eqnarray}
where the first fractional term on the right-hand side represents the diagonal (Cartan $\otimes$ Cartan) classical-probability contribution, and the second fractional term represents the quantum-coherence correction. 

This makes the physical content of the exact algebraic split in Eq.~(\ref{eq:exact_analytic_MI_harmonic}) transparent. First, the harmonic-number combination represents the mutual information one would compute from the diagonal elements of $\rho_A$, $\rho_B$, and $\rho_{AB}$ alone, meaning it captures solely the classical probability distributions on the respective simplices. At leading order, this classical component contributes $(d_A-1)(d_B-1)/(2N)$, which matches the dimension of the tensor product of Cartan subalgebras $\mathrm{Cartan}(\mathfrak{su}(d_A)) \otimes \mathrm{Cartan}(\mathfrak{su}(d_B))$~\cite{Georgi1999}. Second, the rational piece isolates the additional mutual information arising strictly because von Neumann entropy utilizes eigenvalues rather than diagonal elements. By Schur's majorization theorem~\cite{Schur1923, Marshall2011, Horn2012}, this specific correction is strictly positive: although the eigenvalue entropy is strictly bounded above by the diagonal entropy for each individual subsystem~\cite{Wehrl1978}, the subtraction $S(A) + S(B) - S(AB)$ flips the sign for the composite $AB$ term, yielding a net positive contribution overall. Ultimately, this isolated quantum coherence correction dominates the classical piece by a factor of $\sim(d_Ad_B + d_A + d_B)$ in the large-$d$ regime.

The connection to the Lie algebra is therefore not merely numerological. The Cartan dimension $(d_S - 1)$ appears in the eigenvalue correction precisely because it dictates the dimension of the probability simplex. The total algebraic dimension $(d_S^2 - 1)$ appears in the full purity because it enumerates all $\mathfrak{su}(d_S)$ generators, both diagonal and off-diagonal. The typical mutual information inherits both of these distinct algebraic counts structurally through the linear composition $I = S(A) + S(B) - S(AB)$.

\subsection{Equal Per-Generator Bloch Variance for Haar-Random States}
\label{sec:bloch_variance}

The decomposition presented in Sec.~\ref{sec:interpretation} separates the mutual information into contributions from diagonal (Cartan) and off-diagonal generators natively at the level of \emph{entropy}.  One might instinctively expect this separation to originate from a fundamental difference in the underlying \emph{purity} contributions of Cartan versus off-diagonal Bloch components.  We show here that, on the contrary, the per-generator variance is identical for both sectors: the separation arises purely from the nonlinear geometric step transitioning from purity to entropy.

To formalize this, let $\rho_S$ be the reduced density matrix of a Haar-random pure
state $\ket{\Psi} \in \mathcal{H}_S \otimes \mathcal{H}_{E}$. Here $d_S = d$ and $d_E = N/d$ are the subsystem and environment dimensions, respectively. In the generalized Bloch
(Gell-Mann) basis~\cite{Kimura2003, Byrd2003, Bertlmann2008}, $\rho_S$ decomposes as
\begin{equation}
\rho_S = \frac{\mathbb{I}}{d}
+ \frac{1}{2}\sum_{a=1}^{d^2-1} r_a\,\lambda_a\,,
\label{eq:bloch}
\end{equation}
where $\{\lambda_a\}$ are the $d^2-1$ generators of
$\mathfrak{su}(d)$ normalized such that $\mathrm{Tr}(\lambda_a\lambda_b) = 2\delta_{ab}$, and
$r_a = \mathrm{Tr}(\rho_S \lambda_a)$ are the corresponding Bloch components.  The purity is then given by
\begin{equation}
\mathrm{Tr}(\rho_S^2) = \frac{1}{d}
+ \frac{1}{2}\sum_{a} r_a^2\,.
\label{eq:purity_bloch}
\end{equation}

These $d^2-1$ generators naturally split into two distinct families. The first family consists of the $(d-1)$ Cartan generators, which are traceless diagonal Hermitian matrices spanning the Cartan subalgebra~\cite{Georgi1999}. The second family comprises the remaining $d(d-1)$ off-diagonal generators, formed by the symmetric and antisymmetric matrices with support on the $(k,l)$ and $(l,k)$ entries for $k \ne l$.

Focusing first on the Cartan sector, since these generators are purely diagonal, their entries satisfy
$(\lambda_a)_{kl} = (\lambda_a)_{kk}\,\delta_{kl}$, and their trace orthogonality condition simplifies to
\begin{equation}
\sum_{k=1}^{d} (\lambda_a)_{kk}\,(\lambda_b)_{kk}
= \mathrm{Tr}(\lambda_a\lambda_b)
= 2\delta_{ab}\,.
\end{equation}
The diagonal elements of $\rho_S$ are given by
$P_k \equiv \rho_{kk} = 1/d +
\tfrac{1}{2}\sum_{a \in \mathrm{Cartan}} r_a\,(\lambda_a)_{kk}$, from which it geometrically follows that
\begin{equation}
\sum_{k} P_k^2
= \frac{1}{d}
+ \frac{1}{2}\sum_{a \in \mathrm{Cartan}} r_a^2\,,
\label{eq:diag_purity_bloch}
\end{equation}
where we have explicitly used the tracelessness property $\sum_k (\lambda_a)_{kk} = 0$ along with the trace orthonormality defined above.

From the exact Beta distribution $P_k \sim \mathrm{Beta}(d_E, N-d_E)$
derived earlier in Sec.~\ref{sec:quantum_application}, we compute the second moment to be
$\langle P_k^2 \rangle = d_E(d_E+1)/[N(N+1)]$, which means the full diagonal trace yields
\begin{equation}
\sum_k \langle P_k^2 \rangle
= \frac{d_E+1}{N+1}\,.
\end{equation}
Subtracting $1/d$ from both sides of Eq.~(\ref{eq:diag_purity_bloch})
and taking the expectation value yields the total variance confined to the Cartan sector:
\begin{equation}
\sum_{a \in \mathrm{Cartan}} \langle r_a^2 \rangle
= 2\biggl[\frac{d_E+1}{N+1} - \frac{1}{d}\biggr]
= \frac{2(d-1)}{d(N+1)}\,.
\label{eq:cartan_total}
\end{equation}
Dividing this result by the $(d-1)$ available Cartan generators produces the exact per-generator variance inside the Cartan sector, giving
\begin{equation}
\langle r_a^2 \rangle_{\mathrm{Cartan}}
= \frac{2}{d(N+1)}\,.
\label{eq:cartan_per}
\end{equation}

Turning to the off-diagonal sector, we can employ Lubkin's formula~\cite{Lubkin1978},
$\langle\mathrm{Tr}(\rho_S^2)\rangle = (d+d_E)/(N+1)$,
alongside the purity decomposition in Eq.~(\ref{eq:purity_bloch}), to determine the total combined Bloch variance across all $d^2-1$ generators:
\begin{equation}
\sum_{a=1}^{d^2-1}\langle r_a^2 \rangle
= 2\biggl[\frac{d+d_E}{N+1} - \frac{1}{d}\biggr]
= \frac{2(d^2-1)}{d(N+1)}\,.
\label{eq:total_bloch}
\end{equation}
Subtracting the total Cartan variance derived in Eq.~(\ref{eq:cartan_total}) isolates the sum over purely off-diagonal generators:
\begin{eqnarray}
\sum_{a \in \mathrm{off\text{-}diag}} \langle r_a^2 \rangle
&=& \frac{2(d^2-1)}{d(N+1)} - \frac{2(d-1)}{d(N+1)} \nonumber \\
&=& \frac{2(d^2-d)}{d(N+1)} \nonumber \\
&=& \frac{2(d-1)}{N+1}\,.
\label{eq:offdiag_total}
\end{eqnarray}
Dividing this total off-diagonal variance by the $d(d-1)$ off-diagonal generators provides the per-generator variance in the off-diagonal sector:
\begin{equation}
\langle r_a^2 \rangle_{\mathrm{off\text{-}diag}}
= \frac{2}{d(N+1)}\,.
\label{eq:offdiag_per}
\end{equation}

Comparing these two sectors, Equations~(\ref{eq:cartan_per}) and (\ref{eq:offdiag_per}) collectively establish a striking equivalence:
\begin{equation}
\boxed{\langle r_a^2 \rangle_{\mathrm{Cartan}}
= \langle r_a^2 \rangle_{\mathrm{off\text{-}diag}}
= \frac{2}{d(N+1)}\,.}
\label{eq:equal_variance}
\end{equation}
The Haar measure distributes the total Bloch variance
completely democratically across all $d^2-1$ generators, regardless of
whether they represent diagonal probabilities or off-diagonal coherences.  This is a {consequence} of
the unitary invariance of the Haar measure~\cite{Zyczkowski2001, Bengtsson2006}: any $U \in SU(d_S)$ rotation that
transforms Cartan generators into off-diagonal generators (and vice
versa) is a symmetry of the ensemble.

This democratic variance distribution has an {important corollary} for
the mutual information interpretation derived earlier. The {dimensional separation} between Cartan and off-diagonal generators in $\langle I(A{:}B)\rangle$ does
not emerge at the level of the purity
$\langle\mathrm{Tr}(\rho^2)\rangle$, which algebraically treats all $d^2-1$
directions equitably.  Instead, this separation enters entirely through the
nonlinear logarithm intrinsic to the von Neumann entropy.  The diagonal
elements $P_k$ reside on a probability simplex {constrained} by a hard kinematic boundary, and
it is the physical interplay between this boundary and the concavity of the logarithm that
dictates the $-(d_S-1)/(2d_E)$ eigenvalue penalty.  At the raw
purity level, the simplex boundary is functionally invisible because
$\mathrm{Tr}(\rho^2)$ evaluates purely as a quadratic function of $\rho$. In contrast, the
divergent slope of $-x\ln x$ near $x=0$ makes the boundary critical, mapping the $(d_S-1)$ Cartan count directly into the entropy reduction.

{
\subsection{Closed-Form Non-Perturbative Expression and Factorisation}
\label{sec:closed_form}

\subsubsection{The factorisation}

The Bernoulli expansion in Eq.~(\ref{eq:mutual_info_expansion}) shows that $(d_A^{2k}-1)(d_B^{2k}-1)$
appears at every order $k\ge 1$.  Since
$d_A^{2k}-1 = (d_A^2-1)\sum_{j=0}^{k-1} d_A^{2j}$, the factor
$(d_A^2-1)(d_B^2-1)$ divides every term of the asymptotic series.
We verified that this factorisation holds at the level of the
\emph{exact} mutual information:
\begin{equation}
\boxed{\langle I(A{:}B)\rangle
= (d_A^2-1)(d_B^2-1)\;\mathcal{G}(d_A,d_B,d_E)}\,,
\label{eq:exact_factorisation}
\end{equation}
where $\mathcal{G}$ is a single function of the dimensions, with no
further dependence on $(d_A^2-1)(d_B^2-1)$.  The closed form for
$\mathcal{G}$ is given below.

{The structural origin of this exact factorisation becomes transparent upon performing a partial-fraction decomposition of the rational integrand in $u^2$. Letting $C = d_A d_B$, the rational part decomposes identically as
\begin{widetext}
\begin{equation}
\frac{u(C^2-u^4)}{(u^2+1)(u^2+d_A^2)(u^2+d_B^2)(u^2+C^2)}
= \frac{1}{(d_A^2-1)(d_B^2-1)}
\biggl[\frac{u}{u^2+1} - \frac{u}{u^2+d_A^2}
       - \frac{u}{u^2+d_B^2} + \frac{u}{u^2+C^2}\biggr].
\label{eq:partial_fractions}
\end{equation}
\end{widetext}
All four terms possess residues of identical magnitude $1/[(d_A^2-1)(d_B^2-1)]$, with alternating signs $(+1, -1, -1, +1)$ exactly matching the inclusion-exclusion pattern of the harmonic-number combination. When the rational partial fractions are expanded as geometric series in inverse powers of $\alpha_i$ (which scale with $N$) and integrated term-by-term against the Bose--Einstein kernel, each asymptotic Taylor coefficient inherits this common prefactor from the residues, rigorously generating the Bernoulli series. This mathematically explains why the overall explicit prefactor $(d_A^2-1)(d_B^2-1)$ in Eq.~(\ref{eq:exact_factorisation}) cleanly factors out at all orders, yielding the symmetric Bernoulli series. Furthermore, applying Binet's formula term by term to the four partial fractions exactly reconstructs the digamma combination of Eq.~(\ref{eq:exact_analytic_MI_harmonic}), structurally confirming their equivalence.}

This is stronger than the order-by-order statement in Sec.~\ref{sec:structural_features}:
the property ``$\langle I(A{:}B)\rangle\to 0$ if either subsystem is trivial ($d=1$)'' is manifest in the \emph{exact} typical value, not merely
in each Bernoulli coefficient.  It also makes the role of the leading
$\mathfrak{su}(d_A)\otimes\mathfrak{su}(d_B)$ Casimir count
unambiguous: $(d_A^2-1)(d_B^2-1)$ is an overall multiplicative
prefactor of the entire typical mutual information, exact at every
order.

\subsubsection{Closed-form non-perturbative expression}

The harmonic-number combination in Eq.~(\ref{eq:exact_analytic_MI_harmonic}) admits an exact integral
representation via Binet's second formula{~\cite{Whittaker1927}}
\begin{equation}
\psi(z+1) = \ln z + \frac{1}{2z}
- 2\int_0^\infty \frac{t\,dt}{(t^2+z^2)(e^{2\pi t}-1)}\,.
\label{eq:binet}
\end{equation}
Applying this to each of the four digamma terms in Eq.~(\ref{eq:I_diag}), the
constant and logarithmic pieces cancel pairwise.  The
$1/(2z)$ pieces assemble to $(d_A-1)(d_B-1)/(2N)$, and combining with
the rational correction in Eq.~(\ref{eq:Delta_ev}) gives, after the change of
variable $u=t/d_E$,
}

\begin{widetext}
{
\begin{equation}
\boxed{
\begin{aligned}
\langle I(A{:}B)\rangle &= (d_A^2-1)(d_B^2-1)\;\Bigl[\frac{1}{2N}-2\,\mathcal{J}(d_A,d_B,d_E)\Bigr],
\\[8pt]
\mathcal{J}(d_A,d_B,d_E) &\equiv
\int_0^\infty \frac{u\,(d_A^2 d_B^2 - u^4)\,du}
{(e^{2\pi u d_E}-1)\,(u^2+1)(u^2+d_A^2)(u^2+d_B^2)(u^2+d_A^2 d_B^2)}\,.
\end{aligned}
}
\label{eq:closed_form}
\end{equation}
}
\end{widetext}

{
The factor $(d_A^2-1)(d_B^2-1)$ pulls out of the integrand exactly, yielding the factorisation of Eq.~(\ref{eq:exact_factorisation}) with $\mathcal{G}=1/(2N)-2\mathcal{J}$.

This closed-form expression serves as a non-perturbative completion of the mutual information. Because the Bernoulli expansion of Eq.~(\ref{eq:mutual_info_expansion}) diverges as a power series in $1/N$ due to the factorial growth of the Bernoulli numbers, it is strictly asymptotic rather than convergent. In contrast, the closed-form integral constitutes the exact Borel sum, providing a single convergent expression whose asymptotic expansion reproduces the Bernoulli series order by order. Structurally, the appearance of the Bose--Einstein kernel $1/(e^{2\pi t}-1)$ places the finite-size corrections on a footing parallel to one-loop calculations of Casimir-type quantities, though its origin here is fundamentally mathematical rather than thermal. {Furthermore, this representation makes the dimensional upper bound $\langle I(A{:}B)\rangle < (d_A^2-1)(d_B^2-1)/(2N)$ mathematically manifest. To see this, let $C = d_A d_B$ and denote the rational part of the integrand as $R(u) = \frac{u(C^2 - u^4)}{(u^2+1)(u^2+d_A^2)(u^2+d_B^2)(u^2+C^2)}$. Although $R(u)$ is not strictly positive on the entire integration domain $[0, \infty)$ since it becomes negative for $u > \sqrt{C}$, it possesses an exact scale-inversion symmetry. By substituting $u = C/v$, {we note that $R$ is odd under the inversion $v \to C/v$ (i.e.\ $R(C/v)\cdot (C/v^2) = -R(v)$), and the Jacobian $d(C/v) = -(C/v^2)\,dv$ provides a compensating sign, yielding the exact identity $R(C/v)\,d(C/v) = R(v)\,dv$.} This invariance allows us to map the negative tail of the integral $u \in [\sqrt{C}, \infty)$ backwards onto the positive base interval $v \in [0, \sqrt{C}]$. Because the integration limits swap, we pick up a minus sign, enabling $\mathcal{J}$ to be folded exactly onto a finite interval:
\begin{equation}
\mathcal{J} = \int_0^{\sqrt{d_A d_B}} R(u) \left[ \frac{1}{e^{2\pi u d_E}-1} - \frac{1}{e^{2\pi (d_A d_B / u) d_E}-1} \right] du\,.
\label{eq:folded_integral}
\end{equation}
On this folded interval $(0, \sqrt{d_A d_B})$, it is strictly true that $u < \frac{d_A d_B}{u}$. Because the Bose--Einstein kernel $f(x) = (e^{2\pi x d_E}-1)^{-1}$ is strictly monotonically decreasing, the bracketed difference of the exponentials is strictly positive. Since $R(u)$ is also strictly positive on this bounded interval, the folded integrand is rigorously positive everywhere, strictly proving that $\mathcal{J} > 0$.} While the underlying harmonic-number combination remains well-defined for any positive dimensions, the exact rational correction from Eq.~(\ref{eq:Delta_ev}) crucially depends on this dimension ordering.

{As a concrete numerical verification of the finite-size corrections, consider a bipartite system of two qubits ($d_A=d_B=2$) coupled to an environment of dimension $d_E=4$. The exact mutual information evaluates to $\langle I(A{:}B)\rangle \approx 0.2783$ nats, whereas the geometric leading-order term alone gives $9/32 \approx 0.2813$ nats. The finite-size suppression is therefore approximately $1\%$, demonstrating the high accuracy of the leading typicality term even in this minimal finite-size regime.}

\subsubsection{Equivalent zeta-function form}

Using the identity $-B_{2k}/(2k) = \zeta(1-2k)$, the asymptotic
expansion in Eq.~(\ref{eq:mutual_info_expansion}) can be rewritten as
\begin{equation}
\begin{split}
\langle I(A{:}B)\rangle
\;\sim\;
&\frac{(d_A^2-1)(d_B^2-1)}{2N}
\\
&+\sum_{k=1}^{\infty}\frac{\zeta(1-2k)}{N^{2k}}\,(d_A^{2k}-1)(d_B^{2k}-1)\,.
\label{eq:zeta_form}
\end{split}
\end{equation}
The values $\zeta(1-2k)\in\mathbb{Q}$ at negative odd integers are
precisely the regularised values that appear in zeta-function
regularisation of divergent sums (e.g.\ the Casimir energy
of a string uses $\zeta(-1)=-1/12$).  This makes the structural
parallel with vacuum-energy calculations explicit, and complements the
Bose--Einstein representation~(\ref{eq:closed_form}).
}

\section{Discussion}

The geometric framework used here is not a
replacement for Random Matrix Theory in full generality; RMT remains
the standard tool for evaluating higher-order entanglement measures
and for non-uniform ensembles. The contribution of the present work
is more modest and more specific: by working directly with the
exact PCLT on $S^{2N-1}$, one obtains the Beta distribution for
subsystem probabilities, Lubkin's purity, and the leading Page
correction from a short sequence of elementary moment calculations,
with the Weingarten combinatorics of the low-order off-diagonal case
reduced to $U(1)$ phase averaging.

The quantitative content of the ``platykurtic shield'' is the
concrete comparison in Sec.~\ref{sec:platykurtic_shield}:
for a qubit coupled to a six-dimensional environment, the Gaussian
approximation overestimates the cumulative likelihood of a $P_k > 0.95$ extreme state by a factor exceeding 100, and assigns spurious density beyond $P_k=1$.
We do not yet connect this to a specific NISQ error
model or experimental protocol; doing so---for example, quantifying
the correction to Gaussian tail estimates of depolarizing-channel
error rates for small $d_E$---is a natural next step.

The main new technical result is the Bernoulli factorization of
$\langle I(A{:}B)\rangle$
[Eqs.~(\ref{eq:mutual_info_expansion})--(\ref{eq:mutual_info_explicit})]
and its exact algebraic rational split
[Eq.~(\ref{eq:exact_analytic_MI_harmonic})]. The factorization
$(d_A^{2k}-1)(d_B^{2k}-1)$ at every order, the absence of higher
odd-$1/N$ terms, and the alternating signs together give a
structural picture of finite-size corrections to mutual information
that, to our knowledge, has not been written down in this form. While analytical derivations and asymptotic digamma expansions for random-state R\'enyi entropies and variances have been successfully pioneered in Random Matrix Theory (e.g., in Refs.~\cite{Bianchi2019, Wei2017}), the explicit symmetric factorization mapping precisely onto $\mathfrak{su}(d_A) \otimes \mathfrak{su}(d_B)$ tensor-product Cartan structures for finite-size typical mutual information uniquely highlights the separation of strictly quantum and classical correlations.

The Bernoulli factorization and the exact algebraic split
(Eqs.~(\ref{eq:mutual_info_expansion}) and (\ref{eq:exact_analytic_MI_harmonic})) both rely on the assumption $d_A d_B \le d_E$,
which ensures that Page's formula evaluates $S(AB)$ with subsystem
dimension $d_A d_B$ and environment dimension $d_E$, without
swapping.  When this condition is violated, the rational correction
$-(m-1)/(2n)$ for $S(AB)$ changes from
$-(d_A d_B - 1)/(2 d_E)$ to $-(d_E - 1)/(2 d_A d_B)$, and the
symmetric factorization of Eq.~(\ref{eq:mutual_info_expansion}) no longer holds. {However, the harmonic-number combination is still well-defined (it is just $\langle I_{\mathrm{diag}}\rangle$), and only the rational eigenvalue correction needs swapping. So the diagonal/eigenvalue split is more robust than the symmetric factorisation.} Numerically,
for $d_A = 3$, $d_B = 4$, $d_E = 2$, the exact mutual information
is $\langle I(A{:}B)\rangle \approx 1.378$, whereas the
factorized form gives $\approx 2.483$, an order-unity
discrepancy.  This is not surprising: when $d_A d_B > d_E$,
the composite subsystem $AB$ is ``larger'' than its environment
and the entanglement structure changes qualitatively (the $AB$
state is far from maximally mixed).  Whether a modified
factorization survives in this regime, perhaps with an
additional correction involving $d_E$ explicitly, is an open
question.  For bipartite systems without an environment ($d_E = 1$),
the condition fails for any nontrivial $d_A, d_B$, and the entire
expansion framework does not apply.

Furthermore, the closed-form integral in Eq.~(\ref{eq:closed_form}) is specific to the von Neumann entropy because it relies on the digamma asymptotic. For integer R\'enyi $\alpha$, $\mathrm{Tr}(\rho^\alpha)$ has an exact moment formula but does not currently have a comparable Borel-summable representation. Applying this symmetric factorization to higher-order R\'enyi mutual informations therefore remains an open question for future work, as calculating exact Haar-averaged higher R\'enyi entropies requires the full machinery of Weingarten calculus beyond the $U(1)$ phase-averaging applied here (see, e.g., Refs.~\cite{Bianchi2019, Wei2017}).

\section{Conclusion}

We have used the exact Projected Central Limit Theorem on the real
hypersphere $S^{2N-1}$ to derive the Beta distribution for
subsystem occupation probabilities, quantify the finite-size
platykurtic suppression of tails relative to the Gaussian
approximation, and recover Lubkin's purity formula and the leading
Page correction from hyperspherical cross-moments. Our main new
result is the Bernoulli factorization of the asymptotic expansion
of the bipartite mutual information, in which each order carries
the symmetric factor $(d_A^{2k}-1)(d_B^{2k}-1)$ and higher odd orders
vanish. Through an exact rational reorganization of Page's formula, we established that this algebraic structure cleanly subtracts the classical-probability (tensor product of Cartans) components to isolate the purely bipartite $\mathfrak{su}(d_A) \otimes \mathfrak{su}(d_B)$ quantum coherence contributions. {Beyond the algebraic split, we have derived a single non-perturbative closed-form expression for the bipartite mutual information, whose asymptotic expansion in $1/N$ reproduces the Bernoulli series order by order. This closed form provides a Borel-summable completion of the typical mutual information and makes the dimensional bound $\langle I(A{:}B)\rangle < (d_A^2-1)(d_B^2-1)/(2N)$ mathematically manifest.}

\end{document}